# About quantum mechanics interpretation II

**Alexander G. Kyriakos**

*Saint-Petersburg State Institute of Technology,
St.Petersburg, Russia*

Present address:
   *Athens, Greece*
e-mail: lelekous@otenet.gr

## Abstract

In the previous paper [1] we proved that the quantum mechanics, based on the Dirac electron theory, not only has statistic interpretation, but also the specific electromagnetic form. Here, from this point of view we show that all the formal particularities of Dirac's equation have also the known electromagnetic sense.

## Contents

# 1. Introduction

The Dirac equation has a lot of particularities. In modern interpretation these particularities are considered as entirely mathematical features that don't have a physical meaning. For example, according to [2] (section 34-4) "one can prove that all the physical consequences of Dirac's equation do not depend on the special choice of Dirac's matrices. They would be the same if a different set of four 4x4 matrices with the specification,

$$\hat{\alpha}_\mu \cdot \hat{\alpha}_\nu = \begin{cases} 1, & if \quad \mu = \nu \\ 0, & if \quad \mu \neq \nu \end{cases}$$

had been chosen. In particular it is possible to interchange the roles of the four matrices by unitary transformation. So, their differences are only apparent".

The mathematical properties of Dirac's matrices are well known: they are anticommutative and hermitian; they build a group of 16 matrices; the bilinear forms of these matrices have definite transformation properties, which correspond to the vector properties of the (e.g.) electrodynamics; repeatedly it was pointed that in the classical physics these matrices describe the rotations, etc.[3].

We have shown [1] that *the Dirac matrices describe the electromagnetic vectors moving along the curvilinear trajectories (not vectors in the curvilinear space!).* From this suggestion follows the interpretation of all matrix properties.

Below we will consider how the Dirac matrices are joined with the electromagnetic Dirac's equation forms. We will show that all the electron Dirac's equation particularities have the exact electrodynamics meaning. We will also answer to the question, why the "choice of Dirac's matrices" exists, but the results "don't depend on the special choice". Here we will also analyse the physical sense of the different form of Dirac's equation in regard to various Dirac's equation matrix representation.

## 2. Physical sense of the different forms of Dirac's equation

As it is known, there are two Dirac's equation forms

$$\left[ \left( \hat{\alpha}_o \hat{\varepsilon} + c \hat{\vec{\alpha}} \; \hat{\vec{p}} \right) + \hat{\beta} \, mc^2 \right] \psi = 0 , \qquad (2.1)$$

$$\psi^+ \left[ \left( \hat{\alpha}_o \hat{\varepsilon} - c \hat{\vec{\alpha}} \; \hat{\vec{p}} \right) - \hat{\beta} \, mc^2 \right] = 0 , \qquad (2.2)$$

which correspond to the two signs of the relativistic energy expression:

$$\varepsilon = \pm \sqrt{c^2 \vec{p}^{\,2} + m^2 c^4} , \qquad (2.3)$$

Here $\hat{\varepsilon} = i\hbar \dfrac{\partial}{\partial t}$, $\hat{\vec{p}} = -i\hbar \vec{\nabla}$ are the operators of energy and momentum, $\varepsilon$, $\vec{p}$ are the electron energy and momentum, $c$ is the light velocity, $m$ is the electron mass, $\psi$ is the wave function named bispinor and $\hat{\vec{\alpha}}$, $\hat{\beta}$ are the following Dirac's matrices:

$$\hat{\alpha}_0 = \begin{pmatrix} 1 & 0 & 0 & 0 \\ 0 & 1 & 0 & 0 \\ 0 & 0 & 1 & 0 \\ 0 & 0 & 0 & 1 \end{pmatrix}, \quad \hat{\alpha}_1 = \begin{pmatrix} 0 & 0 & 0 & 1 \\ 0 & 0 & 1 & 0 \\ 0 & 1 & 0 & 0 \\ 1 & 0 & 0 & 0 \end{pmatrix},$$

$$\hat{\alpha}_2 = \begin{pmatrix} 0 & 0 & 0 & -i \\ 0 & 0 & i & 0 \\ 0 & -i & 0 & 0 \\ i & 0 & 0 & 0 \end{pmatrix}, \quad \hat{\alpha}_3 = \begin{pmatrix} 0 & 0 & 1 & 0 \\ 0 & 0 & 0 & -1 \\ 1 & 0 & 0 & 0 \\ 0 & -1 & 0 & 0 \end{pmatrix},$$

$$\hat{\alpha}_4 \equiv \hat{\beta} = \begin{pmatrix} 1 & 0 & 0 & 0 \\ 0 & 1 & 0 & 0 \\ 0 & 0 & -1 & 0 \\ 0 & 0 & 0 & -1 \end{pmatrix}$$

Also, as it is known that, for each sign of equation (2.3) there are two hermitian-conjugate Dirac's equations.

Here we consider the physical meaning of all these equations.

Let us consider at first two hermitian-conjugate equations, corresponding to the minus sign in (2.3):

$$\left[ \left( \hat{\alpha}_o \hat{\varepsilon} + c\hat{\vec{\alpha}} \; \hat{\vec{p}} \right) + \hat{\beta} \, mc^2 \right] \psi = 0 , \qquad (2.4)$$

$$\psi^+ \left[ \left( \hat{\alpha}_o \hat{\varepsilon} + c\hat{\vec{\alpha}} \; \hat{\vec{p}} \right) + \hat{\beta} \, mc^2 \right] = 0 , \qquad (2.5)$$

Put, e.g., $\psi = \psi(y)$ and choose

$$\psi = \begin{pmatrix} \psi_1 \\ \psi_2 \\ \psi_3 \\ \psi_4 \end{pmatrix} = \begin{pmatrix} E_x \\ E_z \\ iH_x \\ iH_z \end{pmatrix} \qquad (2.6)$$

Then

$$\psi^+ = \begin{pmatrix} E_x & E_z & -iH_x & -iH_z \end{pmatrix}, \qquad (2.7)$$

Using (2.6) and (2.7) from (2.4) and (2.5) we obtain:

$$\begin{cases} \dfrac{1}{c} \dfrac{\partial E_x}{\partial t} - \dfrac{\partial H}{\partial y} + i\dfrac{\omega}{c} E_x = 0, \\[4pt] \dfrac{1}{c} \dfrac{\partial E_z}{\partial t} + \dfrac{\partial H_x}{\partial y} + i\dfrac{\omega}{c} E_z = 0, \\[4pt] \dfrac{1}{c} \dfrac{\partial H_x}{\partial t} + \dfrac{\partial E_z}{\partial y} - i\dfrac{\omega}{c} H_x = 0, \\[4pt] \dfrac{1}{c} \dfrac{\partial H_z}{\partial t} - \dfrac{\partial E_x}{\partial y} - i\dfrac{\omega}{c} H_z = 0, \end{cases} \qquad (2.8)$$

$$\begin{cases} \dfrac{1}{c}\dfrac{\partial E_x}{\partial t} - \dfrac{\partial H}{\partial y} - i\dfrac{\omega}{c}E_x = 0, \\ \dfrac{1}{c}\dfrac{\partial E_z}{\partial t} + \dfrac{\partial H_x}{\partial y} - i\dfrac{\omega}{c}E_z = 0, \\ \dfrac{1}{c}\dfrac{\partial H_x}{\partial t} + \dfrac{\partial E_z}{\partial y} + i\dfrac{\omega}{c}H_x = 0, \\ \dfrac{1}{c}\dfrac{\partial H_z}{\partial t} - \dfrac{\partial E_x}{\partial y} + i\dfrac{\omega}{c}H_z = 0, \end{cases} \qquad (2.9)$$

As we see, the equation (2.8) and (2.9) are the Maxwell equations with complex currents that we name [1] the spinning semiphoton equations. As we see the equations (2.8) and (2.9) are differed by the current directions. We could foresee this result before the calculations, since the functions $\psi^+$ and $\psi$ are differed by the argument signs:

$$\psi^+ = \psi_0 e^{-i\omega t} \quad \text{and} \quad \psi = \psi_0 e^{i\omega t}.$$

Let us compare now the equations corresponding to the both plus and minus signs of (2.3). For the plus sign of (2.2) we have following two equations:

$$\left[ \left( \hat{\alpha}_o \hat{\varepsilon} - c\hat{\vec{\alpha}}\ \hat{\vec{p}} \right) - \hat{\beta}\ mc^2 \right] \psi = 0, \qquad (2.10)$$

$$\psi^+ \left[ \left( \hat{\alpha}_o \hat{\varepsilon} - c\hat{\vec{\alpha}}\ \hat{\vec{p}} \right) - \hat{\beta}\ mc^2 \right] = 0, \qquad (2.11)$$

The electromagnetic form of the equation (2.10) is:

$$\begin{cases} \dfrac{1}{c}\dfrac{\partial E_x}{\partial t} + \dfrac{\partial H}{\partial y} + i\dfrac{\omega}{c}E_x = 0, \\ \dfrac{1}{c}\dfrac{\partial E_z}{\partial t} - \dfrac{\partial H_x}{\partial y} + i\dfrac{\omega}{c}E_z = 0, \\ \dfrac{1}{c}\dfrac{\partial H_x}{\partial t} - \dfrac{\partial E_z}{\partial y} - i\dfrac{\omega}{c}H_x = 0, \\ \dfrac{1}{c}\dfrac{\partial H_z}{\partial t} + \dfrac{\partial E_x}{\partial y} - i\dfrac{\omega}{c}H_z = 0, \end{cases} \qquad (2.12)$$

Obviously, the electromagnetic form of equation (2.11) will have the opposite signs of the currents compared to (2.12).

Comparing (2.12) and (2.8) we see that the equation (2.12) can be considered as Maxwell's equation in the left co-ordinate system. So as not to use the left co-ordinate system, together with the wave function of the electron $\psi_{ele}$ we can consider the wave function of positron in the form, which correspond to the right co-ordinate system:

$$\psi_{pos} = \begin{pmatrix} E_x \\ -E_z \\ iH_x \\ -iH_z \end{pmatrix}, \quad (2.13)$$

Then, contrary to the system (2.12) we get the system (2.9). The transformation of the function $\psi_{ele}$ to the function $\psi_{pos}$ is named the charge conjugation operation.

Note that the electron and positron wave functions can be considered as the retarded and advanced waves. So the above result links also with the theory of advanced waves of Wheeler and Feynman [4]. (See also Dirac's work on time-symmetric classical electrodynamics [5] and about this theme Konopinski's book [6].

## 3. Physical sense of the matrix choice

Now we will consider the bispinor form of Dirac's equation from the point of view that it describes the spinning object.

Consider the Dirac bispinor equation (2.1)-(2.2) with $\alpha$ - set of Dirac's matrices, as above.

As we saw [1] this matrix sequence $(\hat{\alpha}_1, \hat{\alpha}_2, \hat{\alpha}_3)$ agrees to the spinning semi-photon having $y$-direction. But herewith only the $\hat{\alpha}_2$-matrix is "working", and other two matrices don't give the terms of the equation. The verification of this fact is the Poynting vector calculation: the bilinear forms of $\hat{\alpha}_1, \hat{\alpha}_3$-matrices are equal to zero, and only the matrix $\hat{\alpha}_2$ gives the right non-zero component of Poynting's vector.

The question arises how to describe the photons having initially $x$ and $z$ - directions? It is not difficult to see [2] that the matrices' sequence is not determined by the some special requirements. In fact, this matrices' sequence can be changed without breaking some quantum electrodynamics results.

So we can write three groups of matrices, each of which corresponds to the one and only one direction, introducing the axes' indexes, which indicate the photon direction:

$$(\hat{\alpha}_{1x}, \hat{\alpha}_{2y}, \alpha_{3z}), \quad (\hat{\alpha}_{2x}, \hat{\alpha}_{3y}, \hat{\alpha}_{1z},), \quad (\hat{\alpha}_{2z}, \hat{\alpha}_{1y}, \hat{\alpha}_{3x}).$$

Let us choose now the wave function forms, which give the correct Maxwell equations. We will take as initial the form for the $y$ - direction, which we already used [1], and from them, by means of the indexes' transposition around the circle (see fig.1) we will get other forms for $x$ and $y$ - directions.

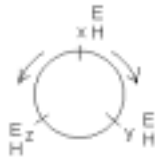

Fig.1

Since in this case the Poynting vector has the minus sign, we can suppose that the transposition takes place counterclockwise.

Let us check the Poynting vector values:

1) $\psi = \psi(y)$, $(\hat{\alpha}_{1_x}, \hat{\alpha}_{2_y}, \hat{\alpha}_{3_z})$, $\psi = \begin{pmatrix} E_x \\ E_z \\ iH_x \\ iH_z \end{pmatrix}$, $\psi^+ = (E_x \ E_z \ -iH_x \ -iH_z)$. (3.1)

$$\psi^+ \hat{\alpha}_{1_x} \psi = (E_x \ E_z \ -iH_x \ -iH_z) \begin{pmatrix} iH_z \\ iH_x \\ E_z \\ E_x \end{pmatrix} = iE_x H_z + iE_z H_x - iE_z H_x - iE_x H_z = 0,$$

$$\psi^+ \hat{\alpha}_{2_y} \psi = (E_x \ E_z \ -iH_x \ -iH_z) \begin{pmatrix} H_z \\ -H_x \\ -iE_z \\ iE_x \end{pmatrix} = E_x H_z - E_z H_x - E_z H_x + E_x H_z =$$

$$= -2(E_z H_x - E_x H_z) = -2[\vec{E} \times \vec{H}]_y$$

$$\psi^+ \hat{\alpha}_{3_z} \psi = (E_x \ E_z \ -iH_x \ -iH_z) \begin{pmatrix} iH_x \\ -iH_z \\ E_x \\ -E_z \end{pmatrix} = iE_x H_x - iE_z H_z - iE_x H_x + iE_z H_z = 0.$$

2) $\psi = \psi(x)$, $(\hat{\alpha}_{2_x}, \hat{\alpha}_{3_y}, \hat{\alpha}_{1_z})$, $\psi = \begin{pmatrix} E_z \\ E_y \\ iH_z \\ iH_y \end{pmatrix}$, $\psi^+ = (E_z \ E_y \ -iH_z \ -iH_y)$. (3.2)

$$\psi^+ \hat{\alpha}_{2_x} \psi = (E_z \ E_y \ -iH_z \ -iH_y) \begin{pmatrix} H_y \\ -H_z \\ -iE_y \\ iE_z \end{pmatrix} = E_z H_y - E_y H_z - E_y H_z + E_z H_y =$$

$$= -2(E_y H_z - E_z H_y) = -2[\vec{E} \times \vec{H}]_x$$

$\psi^+ \hat{\alpha}_{3_y} \psi = 0$,

$\psi^+ \hat{\alpha}_{1_z} \psi = 0$.

3) $\psi = \psi(z)$, $(\hat{\alpha}_{2_z}, \hat{\alpha}_{1_y}, \hat{\alpha}_{3_x})$, $\psi = \begin{pmatrix} E_y \\ E_x \\ iH_y \\ iH_x \end{pmatrix}$, $\psi^+ = \begin{pmatrix} E_y & E_x & -iH_y & -iH_x \end{pmatrix}$. (3.3)

$\psi^+ \hat{\alpha}_{3_x} \psi = 0$,

$\psi^+ \hat{\alpha}_{1_y} \psi = 0$

$\psi^+ \hat{\alpha}_{2_z} \psi = \begin{pmatrix} E_y & E_x & -iH_y & -iH_x \end{pmatrix} \begin{pmatrix} H_x \\ -H_y \\ -iE_x \\ iE_y \end{pmatrix} = E_y H_x - E_x H_y - E_x H_y + E_y H_x =$

$= -2(E_x H_y - E_y H_x) = -2[\vec{E} \times \vec{H}]_z$

As we see, we took the correct result: by the counterclockwise indexes' transposition the wave functions describe the photons, which are moved in negative directions of the corresponding co-ordinate axes.

We may hope that by the clockwise indexes' transposition, the wave functions will describe the photons, which are moved in positive directions of co-ordinate axes. Prove this:

1) $\psi = \psi(y)$, $(\hat{\alpha}_{1_x}, \hat{\alpha}_{2_y}, \alpha_{3_z})$, $\psi = \begin{pmatrix} E_z \\ E_x \\ iH_z \\ iH_x \end{pmatrix}$, $\psi^+ = \begin{pmatrix} E_z & E_x & -iH_z & -iH_x \end{pmatrix}$. (3.4)

$\psi^+ \hat{\alpha}_{1_x} \psi = \begin{pmatrix} E_z & E_x & -iH_z & -iH_x \end{pmatrix} \begin{pmatrix} iH_x \\ iH_z \\ E_x \\ E_z \end{pmatrix} = iE_z H_x + iE_x H_z - iE_x H_z - iE_z H_x = 0$,

$\psi^+ \hat{\alpha}_{2_y} \psi = \begin{pmatrix} E_z & E_x & -iH_z & -iH_x \end{pmatrix} \begin{pmatrix} H_x \\ -H_z \\ -iE_x \\ iE_z \end{pmatrix} = E_z H_x - E_x H_z - E_x H_z + E_z H_x =$

$= 2(E_z H_x - E_x H_z) = 2[\vec{E} \times \vec{H}]_y$

$\psi^+ \hat{\alpha}_{3_z} \psi = \begin{pmatrix} E_z & E_x & -iH_z & -iH_x \end{pmatrix} \begin{pmatrix} iH_z \\ -iH_x \\ E_z \\ -E_x \end{pmatrix} = iE_z H_z - iE_x H_x - iE_z H_z + iE_x H_x = 0$.

2) $\psi = \psi(x)$, $(\hat{\alpha}_{2_x}, \hat{\alpha}_{3_y}, \hat{\alpha}_{1_z},)$, $\psi = \begin{pmatrix} E_y \\ E_z \\ iH_y \\ iH_z \end{pmatrix}$, $\psi^+ = \begin{pmatrix} E_y & E_z & -iH_y & -iH_z \end{pmatrix}$. (3.5)

$$\psi^+ \hat{\alpha}_{2_x} \psi = \begin{pmatrix} E_y & E_z & -iH_y & -iH_z \end{pmatrix} \begin{pmatrix} H_z \\ -H_y \\ -iE_z \\ iE_y \end{pmatrix} = E_y H_z - E_z H_y - E_z H_y + E_y H_z =$$

$$= 2(E_y H_z - E_z H_y) = 2[\vec{E} \times \vec{H}]_x$$

$\psi^+ \hat{\alpha}_{3_y} \psi = 0$,

$\psi^+ \hat{\alpha}_{1_z} \psi = 0$.

3) $\psi = \psi(z)$, $(\hat{\alpha}_{2_z}, \hat{\alpha}_{1_y}, \hat{\alpha}_{3_x})$, $\psi = \begin{pmatrix} E_x \\ E_y \\ iH_x \\ iH_y \end{pmatrix}$, $\psi^+ = \begin{pmatrix} E_x & E_y & -iH_x & -iH_y \end{pmatrix}$. (3.6)

$\psi^+ \hat{\alpha}_{3_x} \psi = 0$,

$\psi^+ \hat{\alpha}_{1_y} \psi = 0$

$$\psi^+ \hat{\alpha}_{2_z} \psi = \begin{pmatrix} E_x & E_y & -iH_x & -iH_y \end{pmatrix} \begin{pmatrix} H_y \\ -H_x \\ -iE_y \\ iE_x \end{pmatrix} = E_x H_y - E_y H_x - E_y H_x + E_x H_y =$$

$$= 2(E_x H_y - E_y H_x) = 2[\vec{E} \times \vec{H}]_z$$

As we see, we also get the correct result.

Now we will prove that the above choice of the matrices give the correct electromagnetic equation forms. Using for example the bispinor Dirac's equation (2.10)

$$[(\hat{\alpha}_o \hat{\varepsilon} - c\hat{\vec{\alpha}}\ \hat{\vec{p}}) - \hat{\beta}\ mc^2\ ]\psi = 0$$

and transposing the indexes clockwise we obtain for the positive direction photons the following results:

1) for $x$-direction:

$$\frac{1}{c}\frac{\partial}{\partial t}\begin{pmatrix}E_y\\E_z\\iH_y\\iH_z\end{pmatrix}+\frac{\partial}{\partial x}\begin{pmatrix}H_z\\-H_y\\-iE_z\\iE_y\end{pmatrix}=-i\frac{mc}{\hbar}\begin{pmatrix}E_y\\E_z\\-iH_y\\-iH_z\end{pmatrix}, \quad\text{or}\quad \begin{cases}\frac{1}{c}\frac{\partial E_y}{\partial t}+\left(\frac{\partial H_z}{\partial x}\right)=-i\frac{\omega}{c}E_y, & a^*\\[4pt] \frac{1}{c}\frac{\partial E_z}{\partial t}-\left(\frac{\partial H_y}{\partial x}\right)=-i\frac{\omega}{c}E_z, & a'\\[4pt] \frac{1}{c}\frac{\partial H_y}{\partial t}-\left(\frac{\partial E_z}{\partial x}\right)=i\frac{\omega}{c}H_y, & a''\\[4pt] \frac{1}{c}\frac{\partial H_z}{\partial t}+\left(\frac{\partial E_y}{\partial x}\right)=i\frac{\omega}{c}H_z, & a^{**}\end{cases} \quad (3.7)$$

2) for $y$-direction:

$$\frac{1}{c}\frac{\partial}{\partial t}\begin{pmatrix}E_z\\E_x\\iH_z\\iH_x\end{pmatrix}+\frac{\partial}{\partial y}\begin{pmatrix}H_x\\-H_z\\-iE_x\\iE_z\end{pmatrix}=-i\frac{mc}{\hbar}\begin{pmatrix}E_z\\E_x\\-iH_z\\-iH_x\end{pmatrix}, \quad\text{or}\quad \begin{cases}\frac{1}{c}\frac{\partial E_z}{\partial t}+\left(\frac{\partial H_x}{\partial y}\right)=-i\frac{\omega}{c}E_z, & b^*\\[4pt] \frac{1}{c}\frac{\partial E_x}{\partial t}-\left(\frac{\partial H_z}{\partial y}\right)=-i\frac{\omega}{c}E_x, & b'\\[4pt] \frac{1}{c}\frac{\partial H_z}{\partial t}-\left(\frac{\partial E_x}{\partial y}\right)=i\frac{\omega}{c}H_z, & b''\\[4pt] \frac{1}{c}\frac{\partial H_x}{\partial t}+\left(\frac{\partial E_z}{\partial y}\right)=i\frac{\omega}{c}H_x, & b^{**}\end{cases} \quad (3.8)$$

3) for $z$-direction:

$$\frac{1}{c}\frac{\partial}{\partial t}\begin{pmatrix}E_x\\E_y\\iH_x\\iH_y\end{pmatrix}+\frac{\partial}{\partial z}\begin{pmatrix}H_y\\-H_x\\-iE_y\\iE_x\end{pmatrix}=-i\frac{mc}{\hbar}\begin{pmatrix}E_x\\E_y\\-iH_x\\-iH_y\end{pmatrix}, \quad\text{or}\quad \begin{cases}\frac{1}{c}\frac{\partial E_x}{\partial t}+\left(\frac{\partial H_y}{\partial z}\right)=-i\frac{\omega}{c}E_x, & c^*\\[4pt] \frac{1}{c}\frac{\partial E_y}{\partial t}-\left(\frac{\partial H_x}{\partial z}\right)=-i\frac{\omega}{c}E_y, & c'\\[4pt] \frac{1}{c}\frac{\partial H_x}{\partial t}-\left(\frac{\partial E_y}{\partial z}\right)=i\frac{\omega}{c}H_x, & c''\\[4pt] \frac{1}{c}\frac{\partial H_y}{\partial t}+\left(\frac{\partial E_x}{\partial z}\right)=i\frac{\omega}{c}H_y, & c^{**}\end{cases} \quad (3.9)$$

So we have obtained three equation groups, each of which contains four equations (totally, there are 12 equations), as it is necessary for the description of all wave directions.

In the same way for all the other Dirac's equation forms the analogue results can be obtained.

Obviously, it is possible via canonical transformations to choose the Dirac matrices so that the initial photon could have not only $x$, $y$ or $z$ directions, but any other direction.

Thus the matrices choice changes only the mathematical description, not the physical results. On the other hand, the knowledge of the Dirac matrix electrodynamics properties can be very useful. So if we knew this fact, e.g., in early times of development of the week interaction theory (i.e. the Fermi theory), the 16 Dirac matrix control wouldn't be needed, since only the vector and pseudovector matrices $\hat{\vec{\alpha}}$ and $\hat{\alpha}_5$ give the electrodynamics invariant (see [1]).

## Conclusion

From above it is not difficult to understand why the matrix choice doesn't have an influence on the Dirac equation solution: the electron properties are invariant in point of the direction motion of the initial photon.